\documentclass[english,draft]{llncs} 

\usepackage{epsf}
\usepackage{amsmath}
\usepackage{amssymb}
\usepackage{latexsym}
\usepackage{hhline}
\usepackage{dcolumn}
\usepackage{verbatim}
\usepackage{varioref}
\usepackage[english]{babel}

\newcommand*{\concat}{\mathbin{.}}

\newcommand{\Bool}{\it{Bool}}

\newcommand{\sseq}{\subseteq}

\newcommand{\Integer}{\mathit{Integer}}
\newcommand{\Int}{\mathit{Int}}
\newcommand{\Natural}{\mathit{Natural}}

\newcommand{\Interv}{\mathit{Interv}}

\newcommand{\Set}{\mathit{Set\ }}
\newcommand{\precision}{\mathit{precision}}

\newcommand{\ch}{\mathit{choose}}
\newcommand{\spli}{\mathit{split}}

\newcommand{\filter}{\mathit{filtering}_{\cal D}}

\newcommand{\branch}{\mathit{branch}}
\newcommand{\naive}{\mathit{naive}}
\newcommand{\ff}{\mathit{ff}}
\newcommand{\fcost}{\mathit{fcost}}

\newcommand{\push}{\mathit{push}}
\newcommand{\topp}{\mathit{top}}
\newcommand{\Sol}{\mathit{Sol}}
\newcommand{\FD}{\mathit{FD}}
\newcommand{\CD}{\Delta}
\newcommand{\p}{\mathit{p}}
\newcommand{\X}{\mathit{X}}
\newcommand{\Doms}{\Delta_{\cal D}}
\newcommand{\RI}{\Re{\cal I}} 

\newcommand{\st}{\mathrel{.}}
\newcommand{\itc}{\mathrel{:}}
\newcommand{\setdiff}{\backslash}

\begin{document}
\pagestyle{headings} 
\title{Branching: the Essence of Constraint Solving}
\titlerunning{Branching: the Essence of Constraint Solving}

\setcounter{page}{1} 
\author{Antonio J. Fern\'{a}ndez\inst{1}\thanks{This work was partly
    supported by EPSRC grants GR/L19515 and GR/M05645 and
    by CICYT grant TIC98-0445-C03-03.}
\and Pat Hill\inst{2}} \institute{Departamento de Lenguajes y Ciencias de la 
Computaci\'{o}n, 
       E.T.S.I.I., 29071 Teatinos, M\'{a}laga, Spain email:afdez@lcc.uma.es
\and  School of Computing, University of Leeds, Leeds, LS2~9JT, 
       England email:hill@comp.leeds.ac.uk}
\maketitle 

\begin{abstract}                                        
   This paper focuses on the branching process for solving any constraint 
   satisfaction problem (CSP).
   A parametrised schema is proposed that
   (with suitable instantiations of the parameters) can solve
    CSP's on both finite and infinite domains. 
   The paper presents a formal specification of the schema
   and a statement of a number of 
   interesting properties that, 
   subject to certain conditions, 
   are satisfied by any instances of the schema.
   It is also shown that the operational procedures of many constraint systems 
   (including cooperative systems) satisfy these conditions.
   
   Moreover, the schema is also used to solve the same CSP 
   in different ways  by means of different instantiations of its 
   parameters.
   
\end{abstract} 

\section{Introduction}

To solve a \emph{constraint satisfaction problem} (CSP), we need to
find an assignment of values to the variables such that all
constraints are satisfied.  A CSP can have many solutions; usually
either any one or all of the solutions must be found. However,
sometimes, because of the cost of finding all solutions,
\emph{partial} CSP's are used where the aim is just to find the best
solution within fixed resource bounds. An example of a partial CSP is
a \emph{constraint optimisation problem} (COP) that assigns a value to
each solution and tries to find an optimal solution (with respect to
these values) within a given time frame.

A common method for solving CSP's is to apply \emph{filtering
algorithms} (also called arc consistency algorithms or propagation
algorithms) that remove inconsistent values from the initial domain of
the variables that cannot be part of any solution. The results are
propagated through the whole constraint set and the process is
repeated until a stable set is obtained. However, filtering algorithms
are, often, incomplete in the sense that they are not adequate for
solving a CSP and, as consequence, it is necessary to employ some
additional strategy called \emph{constraint branching} that divides
the variable domains and then continues with the propagation on each
branch independently.

Constraint Solving algorithms have received intense study from many
researchers, although the focus has been on developing new and more
efficient methods to solve classical
CSP's~\cite{freuder+:extracting-CSP-ijcai95,wallace:why-ac3-better-ac4-ijcai93}
and partial
CSP's~\cite{freuder+:partial-cons-satisf-ai92,meseguer+:cs-global-opt-ijcai95}. See
\cite{kumar:algo-csp-survey-aim92,ruttkay:cs-a-survey,smith:cp-tutorial95,hentenryck:cs-combinatorial-cp95}
for more information on constraint solving algorithms
and~\cite{kondrak+:evaluation-back-algo-ai97,nadel:satisf-algorithms-ci89}
for selected comparisons.

To our knowledge,
despite the fact that it is well known that branching step 
is a crucial process in complete constraint solving,
papers concerned with the general principles of constraint
solving algorithms have mainly focused on the filtering
step~\cite{apt:essence-tcs99,fernandez+:interval-based-flops99,hentenryck+:generic-arc-cons-ai92}.

In this paper, we propose a schema for constraint solving
for both classical and partial CSP's that
includes a generic formulation of the branching process.
(This schema may be viewed as a generalisation and extension of
 the interval lattice-based constraint-solving framework
in~\cite{fernandez+:interval-based-flops99}.)
The schema can be used for most existing
constraint domains (finite or continuous) and, as for the framework
in~\cite{fernandez+:interval-based-flops99}, is also applicable to
multiple domains and cooperative systems.
We will show that the operational procedures of many constraint systems
(including cooperative systems) satisfy these conditions.

The paper is organised as follows. Section~\ref{sect:basic notions} shows the basic 
notions used in the paper and Section~\ref{sect:branching process} describes the 
main functions involved in constraint solving with special attention to those 
involved in the branching step. In Section~ \ref{sect:generic branching} a generic 
schema for classical constraint solving is developed and its main properties are 
declared. Then, Section~\ref{sect:extended schema} extends the original schema for 
partial constraint solving and more properties are declared. 
Section~\ref{sect:instances} shows several instances of the schema to solve both 
different CSP's and different solvings for the same CSP. Section~\ref{sect:conclude} 
contains concluding remarks. Proofs of the properties are found in the Appendixes.

\section{Basic concepts}
\label{sect:basic notions}

Let $D,D_1,\dots,D_n$ be sets or \emph{domains}. Then $\#D$ denotes the cardinality 
of $D$, $\wp(D)$ its power set and $D_<$ denote any totally ordered domain. $\bot_D$ 
and $\top_D$ denote respectively, if they exist, the bottom and top element of $D$ 
and fictitious bottom and top elements otherwise. Throughout the paper, $\CD$ 
denotes a set of domains called \emph{computation domains}.

\begin{definition} (Constraint satisfaction problem)
A \emph{Constraint satisfaction problem} (CSP) is a tuple 
$\langle {\cal V}, {\cal D}, {\cal C} \rangle$ where 
\begin{itemize}
\item  ${\cal V}=\{v_1,\ldots,v_n\}$ is a non-empty finite set of variables.
\item  ${\cal D} = \wp(D_1)\times\ldots\times\wp(D_n)$ where 
       $D_i \in \CD$.
\item  ${\cal C} \subseteq \wp(D_1,\ldots,D_n)$ is a set of 
       \emph{constraints for ${\cal D}$}. 
\end{itemize}
\end{definition} 
  
If, as in the above definition, ${\cal D} = \wp(D_1)\times\ldots\times\wp(D_n)$, 
where $D_i \in \CD$ for all $i \in \{1,\ldots,n\}$, then the set of all constraints 
for ${\cal D}$ is denoted as ${\cal C}_{\cal D}$
and the set $\{D_i \mid 1 \leq i \leq n\}$ is denoted as $\Doms$. 
            
\begin{definition} (Constraint store)
\label{ordering on stores} \label{solution} 
Let $S=(d_1,\ldots,d_n) \in {\cal D}$.
Then $S$ is called a \emph{constraint store for 
$\langle {\cal V}, {\cal D}, {\cal C} \rangle$}. 
$S$ is \emph{consistent} if,  for all
$i \in \{1,\ldots,n\}$, $d_i \neq \emptyset$. 
$S$ is \emph{divisible} if $S$ is consistent and for some 
$i \in \{1,\ldots,n\}$, $\#d_i > 1$. 
Let $S'=(d_1',\ldots,d_n')$ be another constraint 
store for $\langle {\cal V}, {\cal D}, {\cal C} \rangle$. 
Then $S \preceq_s S'$
if and only if $d_i \subseteq d_i'$ for $1 \leq i \leq n$. 

$S$ is a \emph{solution} for 
$\langle {\cal V}, {\cal D}, {\cal C} \rangle$ if 
$S = (\{s_1\},\ldots,\{s_n\})$ and $(s_1,\ldots,s_n) \in c$, 
for all $c \in {\cal C}$.  $S'$ is a \emph{partial solution} for
$\langle {\cal V}, {\cal D}, {\cal C} \rangle$ if there exists a
solution $S''$ for $\langle {\cal V}, {\cal D}, {\cal C} \rangle$ such
that $S'' \prec_s S'$. 
In this case we say that $S'$ \emph{covers} $S''$.
\end{definition}

The set of all solutions for $\langle {\cal V}, {\cal D}, {\cal C} \rangle$ is 
denoted as $\Sol(\langle {\cal V}, {\cal D}, {\cal C} \rangle)$. Note that, if $S 
\in \Sol(\langle {\cal V}, {\cal D}, {\cal C} \rangle)$, then $S$ is consistent and 
not divisible. If $(d_1,\ldots,d_n) \in {\cal D}$ and $i \in \{1\ldots, n\}$, then 
\( 
  (d_1,\ldots,d_n) [d_i/d'] = (d_1,\ldots,d_{i-1},d',d_{i+1},\ldots,d_n).
\)

\begin{example}
Let 
${\cal D} =\wp(\Bool) \times \wp(\Bool) \times \wp(\Bool)$.
Let $c= x \vee (y\wedge z)$ be a constraint for ${\cal D}$. Then
\(
      c = \{ (0,1,1), (1,1,1), (1,0,1), (1,1,0), (1,0,0)\}.
\)
Let 
\begin{align*}
     S_1 &= (\{1\},\{0\},\{0\}),
     &S_2 &= (\{0,1\},\{0,1\},\{0,1\}),\\
     S_3 &= (\{0\},\{0\},\{0\}),
     &S_4 &= (\emptyset,\emptyset,\{0\}).
\end{align*}
Then, $S_1$ is a solution but $S_2$, $S_3$ and $S_4$ are not. 
Note also that $S_1$, $S_2$ and $S_3$
are consistent and $S_4$ is inconsistent.
\end{example}
 
\begin{definition} (Stacks)
\label{ordering on stacks} 
Let $P=(S_1,\ldots,S_\ell) \in \wp({\cal D})$. 
Then $P$ is a {\em stack for} $\langle {\cal V}, {\cal D}, {\cal C} \rangle$.

Let $P'=(S_1',\ldots,S_{\ell'}')$ be another stack for 
$\langle {\cal V}, {\cal D}, {\cal C} \rangle$. 
Then $P \preceq_p P'$ if and only if for
all $S_i \in P$ ($1\leq i\leq \ell$), 
there exists $S_j' \in P'$ ($1\leq j\leq \ell'$) such that 
$S_i \preceq_s S_j'$. 
In this case we say that $P'$ \emph{covers} $P$.
\end{definition}

\section{The Branching Process}
\label{sect:branching process}

This section describes the main functions used in the branching process. 

First we define a filtering function which removes inconsistent values 
from the domains of a constraint store.
\begin{definition} (Filtering function)
\label{filtering function} $\filter::\wp({\cal C}_{\cal D}) \times {\cal D} 
\rightarrow {\cal D}$ is a called a \emph{filtering function for ${\cal D}$} if, for 
all $S \in {\cal D}$, 
\begin{itemize}
\item[(a)]
$\filter({\cal C},S) \preceq_s S$;
  
\item[(b)] 
\(
  \forall R \in \Sol(\langle {\cal V}, {\cal D}, {\cal C} \rangle) \itc
    R \preceq_s S \implies  R \preceq_s \filter({\cal C},S).
\)

\item[(c)] 

If $\filter({\cal C},S)$ is consistent and not divisible then 
   $\filter({\cal C},S)$ is a solution for $\langle {\cal V}, {\cal D}, {\cal C} \rangle$.
\end{itemize}
\end{definition} 

Condition (a) ensures that the filtering never gains values, condition (b) 
guarantees that no solution covered by a constraint store is lost in the filtering 
process and condition (c) guarantees the correctness of the filtering function. 

Variable ordering is an important step in constraint branching.  
We define a \emph{selecting function} which provides a schematic
heuristic for variable ordering.

\begin{definition} (Selecting function)
\label{def:selecting function} Let $S=(d_1,\ldots,d_n) \in {\cal D}$. 
Then 
\[
\ch::\{S \in {\cal D}\mid S \text{ is divisible}\} 
   \rightarrow \{\wp(D) \mid D \in \Doms\} 
\]
is called a \emph{selecting function for} ${\cal D}$ if 
   $\ch(S)=d_j$ where $1 \leq j \leq n$ and $\#d_j > 1$.
\end{definition}

\begin{example}
\label{ex:naive labeling} 
Here is a naive strategy to select the left-most divisible domain.
\begin{align*}
& {\mathbf Precondition:}\ 
  \{S=(d_1,\ldots,d_n) \in {\cal D}
  \text{ is divisible}\} \\ 
& \qquad \ch_{\naive}(S)=d \\ & {\mathbf Postcondition:}\ \{\exists j \in 
\{1,\ldots,n\}\st d=d_j\,,\ \#d_j >1\text{ and } \\ &\qquad\qquad\qquad\qquad  \ 
\forall i \in \{1,\ldots,j-1\} \itc \#d_i = 1 \}. 
\end{align*}  
\end{example}

In the process of branching, 
some computation domain has to be partitioned, 
in two or more parts, in order to introduce a choice point.  
We define a \emph{splitting 
function} which provides a heuristic for value ordering. 

\begin{definition} (Splitting function)
\label{def:splitting function} Let $D \in \CD$ and $k>1$. Then 
\[
\spli_D::\wp{(D)} \rightarrow \underbrace{\wp{(D)} \times \ldots \times 
\wp{(D)}}_{k\ times} 
\]
is called a \emph{splitting function for}
 $D$ if, for all $d \in \wp{(D)}$, $\#d >1$, 
this function is defined $\spli_D(d)=(d_1,\ldots,d_k)$ such that the following 
properties hold: 
\begin{align*}
\text{Completeness}:    &\quad d_1 \cup \ldots \cup d_k = d. \\ 
\text{Contractance}:    
&\quad d_i \subset d,\ \forall i \in \{1,\ldots,k\}. 
\end{align*}
\end{definition}

To guarantee termination, even on continuous domains, an extension of the concept of 
precision map shown in~\cite{fernandez+:interval-based-flops99} is applied here. 

\begin{definition} (Precision map)
\label{precision on stores} Let $\RI = (\Re^+,\Integer)$ 
 where $\Re^+$ is the domain of non-negative reals. 
Then $\precision_D$ is a \emph{precision map} for $D \in \CD$, if  $\precision_D$ is 
a strict monotonic function  from  $\wp(D)$ to $\RI$. 

Let $S=(d_1,\ldots,d_n)$ be a constraint store for 
$\langle {\cal V}, {\cal D}, {\cal C} \rangle$ and, for each
$D \in \Doms$, $\precision_{D}$ is defined for $D$.
Then, a {\em precision map for ${\cal D} = (D_1,\dots,D_n)$} is defined as
\begin{align*}
\precision(S)=\sum_{1 \leq i \leq n} \precision_{D_i}(d_i), 
\end{align*}
where the sum in $\RI$ is defined as $(a_1,a_2)+(b_1,b_2)=(a_1+b_1,a_2+b_2)$. 
\end{definition}

The monotonicity of the precision is a direct consequence of the 
definition\footnote{%
$\RI$ is continuous and infinite so that it is supposed that we can define a map 
from $D$ to $\RI$, even if $D$ is infinite. $\RI$ was chosen since it is valid for 
the interval domain as it is shown in Section~\ref{sect:instances}, but any  domain 
totally ordered supporting the operator $-$ may also be adequate (see 
Line~\ref{lab:main3} in Figure~\ref{fig:generic schema}).}. 

\begin{proposition}
\label{precision monotonicity} Let $S,S'$ be two constraint stores for $\langle 
{\cal V}, {\cal D}, {\cal C} \rangle$. If $S \prec_s S'$ then 
 $\precision(S) <_{\RI}$ $\precision(S')$. 
\end{proposition}

The precision map also means a novel way to 
normalise the selecting functions when 
the constraint system supports multiple domains. 

\begin{example}
\label{ex:first fail}
 The well known \emph{first fail principle} chooses the 
variable constrained with the smallest domain. For multiple domain constraint 
systems 
to emulate the first fail principle, 
we define $\ch/1$ so that it selects the domain 
with the smallest precision\footnote{%
It is straightforward to include more conditions e.g., if
$d_i,d_k,d_j$ have the same (minimum) precision, the most left domain
can be chosen i.e., $d_{minimum(i,k,j)}$.}. 
We denote this procedure by $\ch_{\ff}$.

\begin{align*}
& {\mathbf Precondition:}\ 
  \{S=(d_1,\ldots,d_n) \in {\cal D}
  \text{ is divisible}\} \\ 
& \qquad \ch_{\ff}(S)=d \\ 
& {\mathbf Postcondition:}\ 
  \{\exists j \in \{1,\ldots,n\}\st d=d_j\,, \#d_j>1\text{ and }\\ 
& \quad \forall i \in \{1,\ldots,n\}\backslash 
  \{j\} \itc \#d_i>1 \implies 
    \precision_{D_j}(d_j) \leq_{\RI} \precision_{D_i}(d_i) \}.
\end{align*}                                             
\end{example}

\section{Branching in Constraint Solving} 
\label{sect:generic branching} 

Figure~\ref{fig:generic schema} shows a generic schema for solving any CSP $\langle 
{\cal V}, {\cal D}, {\cal C} \rangle$. This schema  requires the following 
parameters: ${\cal C}$, the set of constraints to solve, a constraint store $S$ for 
$\langle {\cal V}, {\cal D}, {\cal C} \rangle$, a bound $p \in \RI$ and a 
non-negative real bound $\varepsilon$. There are a number of values and subsidiary 
procedures that are assumed to be defined externally to the main branch procedure: 
\begin{itemize}
\item
    a filtering function $\filter/2$ for ${\cal D}$;

\item 
    a selecting function $\ch/1$ for ${\cal D}$;
\item 
    a splitting function $\spli_D$ for each
    domain 
    $D \in \Doms$;      
\item 
      a precision map for ${\cal D}$ (therefore it is assumed that  
      there is defined one precision map for each  
      $D \in \Doms$);      
\item  
      a stack $P \in \wp({\cal D})$ for $\langle {\cal V}, {\cal D}, {\cal C} \rangle$.
\end{itemize}
It is assumed that all the external procedures have an implementation that 
terminates for all possible values.         

\begin{figure}[htb]
\begin{align}
         & \text{\textit{procedure }} 
     \branch({\cal C},S,\p,\varepsilon) \nonumber\\
         & \text{\textit{begin}}  \nonumber\\                        
\tag{1}\label{lab:main1}
         &\quad S \leftarrow \filter({\cal C},S); \\ 
\tag{2}\label{lab:main2}
         &\quad \text{if }S\text{ is consistent then} \\ 
\tag{3}\label{lab:main3}
         &\qquad \text{if (}S \text{ is not divisible}
         \text{ or } \p<\top_{\RI} \text{ and }\p - \precision(S) \leq (\varepsilon,0)
     \text{) then} \\ 
\tag{4}\label{lab:main4} 
         &\qquad\qquad\push(P,S);
   \qquad\qquad\qquad\%\%
     \text{ Add $S$ to top of $P$} \nonumber \\ 
\tag{5}\label{lab:main5}
         &\qquad \text{else } \nonumber\\ 
\tag{6}\label{lab:main6}
         &\qquad\quad d_j \leftarrow \ch(S); \\ 
\tag{7}\label{lab:main7}
         &\qquad\quad (d_{j1},\ldots,d_{jk})\leftarrow 
\spli_{D_j}(d_j), \text{ where }d_j \subseteq D_j;\\ 
\tag{8}\label{lab:main8} 
         &\qquad\ \ \left.
     \begin{array}{l}    
       \branch(C,S[d_j/d_{j1}],\precision(S),\varepsilon) \quad \vee \\ 
       \quad \ldots \qquad \ldots \qquad \ldots \qquad \ldots\qquad \ldots\quad\vee \\ 
       \branch(C,S[d_j/d_{jk}],\precision(S),\varepsilon);       
     \end{array}
     \right\}\%\%\text{ Choice Points} \\
&\qquad\text{\textit{endif};} \nonumber\\           &\quad\text{\textit{endif};} 
\nonumber\\ & \text{\textit{end.}}  \nonumber 
\end{align} 
\caption{\footnotesize $\branch/4$: A Generic Schema for Constraint Solving} 
\label{fig:generic schema} 
\end{figure}                                                    
\normalsize                                                       

\begin{theorem} (Properties of the $\branch/4$ schema)
\label{properties of the schema} 
Let $S$ be the top element in ${\cal D}$ (i.e., 
$S=(D_1,...,D_n)$), $\varepsilon \in \Re^+$ and $\p = \top_{\RI}$. Then, the 
following properties are guaranteed: 
\begin{enumerate}  
\item \emph{Termination}: if $\varepsilon > 0.0$ then
      $\branch({\cal C},S,\p,\varepsilon)$ terminates;

\item \emph{Completeness}: 
       if $\varepsilon = 0.0$ and the execution of
       $\branch({\cal C},S,\p,\varepsilon)$ terminates, then the final
       state for the stack $P$ contains all the solutions for 
       $\langle {\cal V}, {\cal D}, {\cal C} \rangle$;
\item \emph{Approximate completeness}:
      if $\varepsilon > 0.0$ and $R$ is a  solution
      for $\langle {\cal V}, {\cal D}, {\cal C} \rangle$, 
      then an execution of $\branch({\cal C},S,\p,\varepsilon)$
      will result in $P$ containing either $R$
      or a partial solution $R'$ that covers $R$.
\item \emph{Correctness}:
      if $\varepsilon = 0.0$,
      the stack $P$ is initially empty
      and the execution of $\branch({\cal C},S,\p,\varepsilon)$
      terminates with  $R$ in the final state of $P$,
      then
      $R$ is a solution for $\langle{\cal V}, {\cal D}, {\cal C} \rangle$.
\item \emph{Approximate correctness or control on the result precision}:
      If $P_{0.0}$, $P_{\varepsilon_1}$
      and $P_{\varepsilon_2}$ are stacks resulting  
      from any terminating execution 
      of $\branch({\cal C},S,\p,\varepsilon)$
     (where initially $P$ is empty)
     when $\varepsilon$ has the values $0.0$,
     $\varepsilon_1$ and $\varepsilon_2$,    respectively, 
     $0.0 < \varepsilon_1 < \varepsilon_2$ and $P_{0.0}$ is not empty,
     then
\(
P_{0.0} \preceq_p P_{\varepsilon_1} \preceq_p P_{\varepsilon_2}.
\)

     (In other words, the set of (possibly partial) solutions  in the final state of
     the stack is dependent on the value of $\varepsilon$ in the sense that lower
     $\varepsilon$ is, closer to the real set of solutions is).
\end{enumerate}     
\end{theorem}

Observe that the bound $\varepsilon$ guarantees termination and allows to control 
the precision of the results. 

\section{Solving optimisation problems}
\label{sect:extended schema} 

The schema in Figure~\ref{fig:generic schema} can be adapted to solve COPs 
by means of three new subsidiary functions.
\begin{definition} (Subsidiary functions and values)
\label{def:new-subsidiary-functions}
Let $D_<$ be a totally ordered domain\footnote{%
      Normally $D_<$ would be $\Re$.}.
Then we define
\begin{itemize}
\item  a \emph{cost function},
       $\fcost::{\cal D} \rightarrow D_<$;                
\item 
     an \emph{ordering relation},
     $\diamond::D_< \times D_< \in \{>,<,=\}$;
\item a \emph{bound}, $\delta \in D_<$.                                           
\end{itemize}
\end{definition}

Then the {\em extended schema}, $\branch_+/4$, is 
obtained from the schema $\branch/4$ by replacing 
Line~\ref{lab:main4} in Figure~\ref{fig:generic schema} with: 
\begin{align*}
\tag{4*}\label{lab:main4*} 
         &\text{if }\fcost(S)\diamond \delta \text{ then }
         \delta \leftarrow \fcost(S);\  \push(P,S) \text{ endif;} \nonumber
\end{align*}

\begin{theorem} (Properties of the $\branch_+/4$ schema)
\label{properties of the extended schema} Let $S$ be the top element in ${\cal D}$ 
(i.e., $S=(D_1,...,D_n)$), $\varepsilon \in \Re^+$ and $\p = \top_{\RI}$. Then, the 
following properties hold: 
\begin{enumerate}
\item \emph{Termination}: if $\varepsilon > 0.0$, then
      the execution of
      $\branch_+({\cal C},S,\p,\varepsilon)$ terminates;

\item If $\fcost$ is a constant function with value $\delta$ and
      $\diamond$ is $=$, then all properties shown in
      Theorem~\ref{properties of the schema} hold for the execution of
      $\branch_+({\cal C},S,\p,\varepsilon)$.

\item \emph{Soundness on optimisation}: 
      if $\varepsilon=0.0$, $\diamond$ is $>$ (resp. $<$), 
      $\delta = \bot_{D_<}$ (resp. $\top_{D_<}$),
      the stack $P$ is initially empty and
      the execution of $\branch_+({\cal C},S,\p,\varepsilon)$
      terminates with $P$ non-empty,
      then the top element of $P$ is the first solution found that 
      maximises (resp. minimises) the cost function.
\end{enumerate}
\end{theorem}
        
Unfortunately, if $ \varepsilon$ is higher than 0.0, we cannot
guarantee that the top of the stack contains a solution or even a
partial solution for the optimisation problem. 
However, by imposing a monotonicity condition on the cost function
$\fcost/1$, we can compare solutions.

\begin{theorem}
\label{more properties on optimisation} 
(More properties on optimisation) 
Suppose 
that, for $i \in \{1,2\}$, $P_{\varepsilon_i}$ is a stack resulting  from the 
execution of  $\branch_+({\cal C},S,\p,\varepsilon_i)$ where 
$\varepsilon_i \in \Re^+$. 
Suppose also that $\topp(P)$ returns the top element of a non empty stack 
$P$. Then, if $\varepsilon_1 < \varepsilon_2$ the following property hold. 

    \emph{Approximate soundness}:
      If for $i \in \{1,2\}$, $P_{\varepsilon_i}$ is not empty,
      and $\topp(P_{\varepsilon_2})$ is a solution or covers a solution for 
     $\langle{\cal V}, {\cal D}, {\cal C} \rangle$,            
      then, if $\fcost/1$ is monotone and $\diamond$ is $<$ (i.e., a minimisation problem),
     \begin{align*}
            &     \fcost(\topp(P_{\varepsilon_1})) \preceq_{D_<} 
                 \fcost(\topp(P_{\varepsilon_2})),  \\ 
     \intertext{%
      and, if $\fcost/1$ is anti-monotone and $\diamond$ is $>$ (i.e.,a maximisation problem),
} 
&     \fcost(\topp(P_{\varepsilon_1})) \succeq_{D_<} 
\fcost(\topp(P_{\varepsilon_2})). 
   \end{align*}
\end{theorem}

Therefore, by using a(n) (anti-)monotone cost function, the lower
$\varepsilon$ is, the better the (probable) solution is.  Moreover,
decreasing $\varepsilon$ is a means to discard approximate solutions.
For instance, in a minimisation problem, if
\[
\fcost(\topp(P_{\varepsilon_1})) \succ_{D_<} \fcost(\topp(P_{\varepsilon_2}))
\]
 with $\fcost/1$ monotone, then, by
the {\em approximate soundness} property it is deduced that
$\topp(P_{\varepsilon_2})$ cannot be a solution or cover a solution.
       
\section{Examples}
\label{sect:instances}

To illustrate the schemas  $\branch/4$ and  $\branch_+/4$ presented in the 
previous two sections, 
several instances of  $\branch/4$ are given 
for some well-known domains of computation.
In addition, we explain how the choice of instantiation of 
the additional global functions and parameters in the 
definition of $\branch_+/4$ can determine the method of solution for the CSP. 
 
\subsection{Some instances}

In the following, $\branch_{\X}$ denotes an instance of the schema 
$\branch/4$ for 
solving the CSP $\langle{\cal V}, {\cal D}, {\cal C} \rangle$ where 
$\X \subseteq \Doms$. We assume that 
\begin{align*}
\CD = &\{\Bool, \Integer, \Re, \Set\Integer\} \cup \{\Interv(D)\mid 
   D \text{ is a lattice}\}. 
\end{align*}
where $Interv(D)$ denotes the set 
$\{(d_1,d_2)\mid d_1,d_2 \in D, d_1 \leq d_2\}$. 

To identify $\branch_{{\cal D}}$, we indicate a possible definition for both the 
splitting function and the precision map for each 
$D \in \CD_{\cal D}$ and assume 
that both a  selecting function (e.g., $\ch_{\ff}$ as defined in 
Example~\ref{ex:first fail}) and a filtering function for ${\cal D}$ have been 
already defined. We also  indicate the initial value of $S \in {\cal D}$, so that 
the execution of $\branch_{{\cal D}}({\cal C},S,\p,\varepsilon)$ allows to solve the 
CSP where $\varepsilon \in \Re^+$. 

\subsubsection{The finite domain ($\FD$)} 
Constraint solving in a $\FD$ of sparse elements is solved by an instance
$\branch_\FD$ as defined below where $\spli_\FD$ is defined as a naive enumeration 
strategy in which values are chosen from left to right. For example, consider a 
finite domain of integers \cite{dincbas+:chip-fgcs88}, Booleans 
\cite{codognet+:local-prop-jar96} or finite sets of integers 
\cite{dovier+:log-jlp96}). 
                     
\begin{align*} 
&\FD \in \{\Integer, \Bool, \Set \Integer\},\\
& \branch_\FD 
    \left\{
     \begin{array}{l}
         S=(\underbrace{\FD,\ldots,\FD}_{n\ times}); \\       
        \precision_{\FD}(d)=(\# d,0); \\                 
        \spli_{\FD}(\{a_1,a_2,a_3,\ldots,a_k\})=(\{a_1\},\{a_2,a_3,\ldots,a_k\}).
     \end{array}
     \right.
\end{align*}

\subsubsection{Finite closed intervals} 
\label{close FD interval instances} 

Many existing $\FD$ constraint systems solve constraints defined in
the domain of closed intervals $[a,b]$ where $a,b \in \FD$ and denoted
here by $a..b$. Usually $a,b$ are either
integers~\cite{codognet+:clp(fd)-jlp96}, Booleans\footnote{The Boolean
domain is considered as the integer subset
$\{0,1\}$.}~\cite{codognet+:local-prop-jar96} or finite sets of
integers~\cite{gervet:clp(sets)-constraints97}. Here are two instances
of our schema that solve CSP's on these domains:
\begin{align*} 
&\FD \in \{\Integer,\Bool\},\\
&\branch_{\mathit{\Interv(\FD)}}\  
     \left\{
     \begin{array}{l}
      S=(\underbrace{\bot_{\FD}..\top_{\FD},\ldots,\bot_{\FD}..\top_{\FD}}_{n\ times}); \\      
      \precision_{\Interv(\FD)}(a..b)=(b-a,0); \\      
      \spli_{\Interv(\Integer)}(a..b)=(a..a,a+1..b).
     \end{array}
     \right.
\end{align*}                              

\begin{align*} 
&\FD = \Set \Int,\\
&\branch_{\Interv(\FD)}\  
     \left\{
     \begin{array}{l}
       S=(\underbrace{\emptyset..\Integer,\ldots,\emptyset..\Integer}_{n\ times});  \\            
       \precision_{\Interv(\FD)}(a..b)=(\#b-\#a,0); \\        
       \spli_{\Interv(\FD)}(a..b)=(a..b\backslash\{c\},a \cup\{c\}..b)  
             \text{ where }c \in b\backslash a.
     \end{array}
     \right.
\end{align*}

\subsubsection{Lattice (interval) domain} 
In \cite{fernandez+:interval-based-flops99}, we have described a generic filtering
algorithm that propagates interval constraints on any domain $L$ with lattice 
structure subject to the condition that a function $\circ_L::L \times L \rightarrow 
\Re$ is defined that is strictly monotonic on its first argument and strictly 
anti-monotonic on its second argument.  Below we provide an instance to solve any 
CSP defined on $\Interv(L)$: 

\begin{align*} 
\branch_{\mathit{\Interv(L)}}\  
     \left\{
     \begin{array}{l}
     L \text{ is a lattice} \text{ and }
     S=(\underbrace{[\bot_{L},\top_{L}],\ldots,[\bot_{L},\top_{L}}_{n\ times}]); \\      
       \precision_{\Interv(L)}(r)=  \left\{
                      \begin{array}{c}
                         (b \circ_L a,2) \text{ if } r=[a,b];\\ 
                         (b \circ_L a,1) \text{ if } r=(a,b];\\
                         (b \circ_L a,1) \text{ if } r=[a,b);\\                
                         (b \circ_L a,0) \text{ if } r=(a,b);        
                      \end{array}
                      \right.   \\
      \spli_{\Interv(L)}(\{a,b\})=(\{a,c],(c,b\})\text{ where }a \preceq_L c \prec_L b.
     \end{array}
     \right.
\end{align*}

\noindent $\{a,b\}$ denotes any interval in $L$. 
With this instance we
have a constraint solving mechanism for solving (interval) constraints
defined on any domain with lattice structure. 
Thus it is a good complement to the filtering algorithm in
\cite{fernandez+:interval-based-flops99}.  
Note also that if $L$ is
$\Re$ and $\circ_L$ is $-$, we obtain the instance
$\branch_{\Interv(\Re)}$ (also, if $c=\frac{b-a}{2.0}$ we have a usual
strategy of real interval division at the mid point).

\subsubsection{A cooperative domain} 
The schema also supports cooperative instances
that solve CSP's defined on multiple domains. 
This is done by mixing together several instances of the schema $\branch/4$. 
As an example, consider
$\branch_{\mathit{BNR}}$ as defined below where $\spli_{\Interv(D)}$
and $\precision_{\Interv(D)}$ are defined as in previous examples for 
$D \in \{\Bool, \Integer, \Re\}$:

\begin{align*} 
\branch_{\mathit{BNR}}\  
     \left\{
     \begin{array}{l}
        \CD=\{\Interv(D) \mid D \in \{\Bool, \Integer, \Re\} \}. \\       
        S=(\underbrace{\emptyset..D_1,\ldots,\emptyset..D_n}), 
        \{D_1,\ldots, D_n\}  \sseq \{\Bool,\Integer,\Re\}.
     \end{array}
     \right.               
\end{align*}                 

This instance simulates the well known \emph{splitsolve} method of 
CLP(BNR)~\cite{benhamou:applying-jlp97}. 

The generic schema is also valid for solving non-linear constraints provided the 
filtering function $\filter/2$ solves constraints in non-linear form. 

\subsection{Different ways to solve the instances of a CSP} 

Here we show that, for any instance, the schema $\branch_+/4$ also allows a CSP to 
be solved in many different ways, depending on the values for $\fcost$, $\delta$ and 
$\diamond$. 

For instance, a successful result for a classical CSP can either be all possible 
solutions or a set of partial solutions that cover the actual solutions. 
As stated in 
Theorem~\ref{properties of the extended schema}(2), to solve classical CSP's, 
$\fcost$ should be defined as the constant function $\delta \in \Re$ 
and the parameter 
$\diamond$ should have the value $=$. 
In Table~\ref{CSP type table} this case is given in 
the first row. 

As well, as shown in Theorem \ref{properties of the extended
schema}(3), a CSP is solved as a COP by instantiating $\diamond$ as
either $>$ (for maximisation problems) or $<$ (for minimisation
problems). 
The value $\delta$ should be instantiated to the initial cost value
from which an optimal solution must be found.  
Traditionally, the
range of the cost function (i.e., $D_<$) is the domain $\Re$.  
Rows 2 and 3 in Table~\ref{CSP type table} show how to initialise both
$\delta$ and $\diamond$ in these two cases.

\begin{table}[thb]
\centerline{ 
\begin{tabular}{|c|c|c|c|c|}    \hline
CSP Type                 &       $\fcost$         &      $D_{<}$    &$\ \diamond\ $ 
& $\ \delta\ $   \\ \hline Classical CSP            &      constant          &      
$\Re$       &   $=$         & $\fcost(S)$ \\ Typical Minimisation COP &   any cost 
function    &      $\Re$      &   $<$         & $\top_{\Re}$ \\ Typical Maximisation 
COP &   any cost function    &      $\Re$      &   $>$         & $\bot_{\Re}$ \\  
Max-Min      COP         &   any cost function    & $\Re\times\Re$  &   $<$         
& $(\top_{\Re},\bot_{\Re})$ \\ \hline 
\end{tabular}
}
 
\caption{CSP type depends on parameters instantiation} \label{CSP type table} 
\end{table}
\normalsize

In contrast to typical COP's that maintain either a lower bound or an upper bound, 
our schema also permits a mix of the maximization and minimization criteria (or even 
to give priority to some criteria over others). This is the case when $D_<$ is a 
compound domain. Then the ordering in $D_<$ determines how the COP will be solved. 

\begin{example}
Let $\langle {\cal V}, {\cal D}, {\cal C} \rangle$ be a COP,  $D_<$ the domain 
$\Re^2$ with ordering $(a,b) < (c,d) \iff a < c \wedge b > d$, 
$\fcost(S)=(\fcost_1(S),\fcost_2(S))$ a cost function on $\Re^2$ for any $S \in  
{\cal D}$ where $\fcost_1,\fcost_2::{\cal D}\rightarrow \Re$.  Then, if $\delta$ and 
$\diamond$ are as shown in Row 4 of Table~\ref{CSP type table}, this COP is solved 
by minimising $\fcost_1$ and maximising $\fcost_2$. However, if $<$ is defined 
lexicographically, this COP is solved by giving priority to the minimisation of 
$\fcost_1$ over the minimisation of $\fcost_2$ e.g. suppose $S_1$, $S_2$ and $S_3$ 
are solutions with costs $(1.0,5.0)$, $(3.0,1.0)$ and $(1.0,8.0)$, respectively. 
Then with the first ordering the optimal solution is $S_3$ whereas with the 
lexicographic ordering $S_1$ is the optimal solution). 
\end{example}

\section{Concluding remarks} 
\label{sect:conclude} 

This paper analyses the branching process in constraint solving. 
We have provided a generic schema for solving CSP's 
on finite or continuous domains 
as well on multiple domains.
We have proved key properties such as correctness and completeness. 
We have shown how termination may be guaranteed by means of a 
\emph{precision map}.  
We have also shown, by means of an example, 
how, for systems supporting multiple domains, the precision map can be used to
normalise the heuristic for variable ordering.

By using a schematic formulation for the branching process, we have
indicated which properties of main procedures involved in branching
are responsible for the key properties of constraint solving. 
For optimisation problems, we have also shown by means of examples that,
in some cases, the methods for solving CSP's depend on the ordering of
the range of the cost functions.

By combining a filtering function satisfying our conditions 
with an appropriate
instance of our schema, we obtain an
operational semantics for a constraint programming domain
(for example: FD, sets of integers, Booleans, multiple
domains, ...,etc) and systems designed for constraint solving such as
clp(FD)~\cite{codognet+:clp(fd)-jlp96},
clp(B)~\cite{codognet+:clp(b)-plilp'94},
DecLic~\cite{goualard+:hybrid-jflp99},
clp(B/FD)~\cite{codognet+:local-prop-jar96},
CLIP~\cite{hickey:clip-padl2000},
Conjunto~\cite{gervet:clp(sets)-constraints97} or
CLP(BNR)~\cite{benhamou:applying-jlp97}.

Further work is needed to consider how to construct an efficient 
implementation\footnote{%
For example, in COP problems, the top of $P$ can be
removed after Line~\ref{lab:main4} 
of the schema so that $P$ contains
only the optimal solution found so far and memory can be saved.}.  
Moreover, it would be useful to examine how the
efficiency of a COP solver in our schema could be improved by adding
constraints $\fcost(S)\diamond\delta$ to the original set of
constraints for solving ${\cal C}$, so that exhaustive search is replaced
by a forward checking mechanism.
\bibliographystyle{plain}
\bibliography{biblio}

\section*{Appendix: Proofs}
A path $q \in (\Natural \setdiff \{0\})^*$ is any finite sequence of (non-zero) 
natural numbers.  The empty path is denoted by $\varepsilon$, whereas $q \concat i$ 
denotes the path obtained by concatenating the sequence formed by the natural number 
$i\neq 0$ with the sequence of the path $q$. The length of the sequence $q$ is 
called the \emph{length} of the path $q$. 

Given a tree, we label the nodes by the paths to the nodes. The root node is 
labelled $\epsilon$. If a node with label $q$ has $k$ children, then they are 
labelled, from left to right, $q \concat 1, \ldots, q \concat k$. 

\begin{definition}(Search tree).
Let $S$ be a constraint store for $\langle {\cal V}, {\cal D}, {\cal C} \rangle$, 
$\varepsilon \in \Re$ and $\p \in \RI$. The search tree for $\branch({\cal 
C},S,p,\varepsilon)$ is a tree that has $S$ at the root node and, as children, has 
the search trees for the recursive executions of $\branch/4$ as consequence of 
reaching Line~\ref{lab:main8} of Figure~\ref{fig:generic schema}. 

Given a search tree for $\branch({\cal C},S,p,\varepsilon)$, we say that $S_\epsilon 
= S$ is the constraint store and 
$p_\epsilon=\p$ the precision at the root node 
$\epsilon$. Let $S_q$ be the  constraint store and $\p_q$ the precision at a node 
$q$. If $q$ has $k > 0$ children $q \concat 1, \ldots, q \concat k$, then $S_q$ is 
consistent and, if $S_q^f = \filter({\cal C},S_q)$, then $S_q^f$ is divisible so 
that 
$\ch(S^f_q) = d_j$ and, for some $k > 0$, 
 $\spli_D(d_j) = (d_{j1},\ldots,d_{jk})$.
Then we say that $S_{q \concat i} = S^f_q[d_j/d_{ji}]$ is the constraint store and 
$p_{q \concat i} = \precision(S^f_q)$ the precision at node $q \concat i$, for $i 
\in \{1,\ldots,k\}$. 
\end{definition}

\begin{lemma}
\label{lemma:ordering on splitting} Let $\ch/1$ be a selecting function for ${\cal 
D}$, $\spli_D/1$ a splitting function for $D \in \Doms$, $S=(d_1,\ldots,d_n)$ a 
consistent and divisible constraint store for $\langle {\cal V}, {\cal D}, {\cal C} 
\rangle$, $d_j = \ch(S)$, $d_j \subseteq D$ and $(d_{j1},\ldots,d_{jk}) = 
\spli_{D}(d_j)$. Then 
\begin{itemize}
\item[(a)]

       $\forall i \in \{1,\ldots,k\}: S[d_j/d_{ji}]\prec_s S$.
       
\item[(b)] 

       Also, if $S'$ is a solution for 
       $\langle {\cal V}, {\cal D}, {\cal C} \rangle$ and 
       $S' \prec_s S$, then 
       \[
       \exists i \in \{1,\ldots,k\}: S' \preceq_s S[d_j/d_{ji}].
       \]
\end{itemize}
\end{lemma}
\begin{proof} We prove the cases separately.

\textbf{Case (a).} 
   By Definition~\ref{def:selecting function}, 
   $\#d_j>1$ and, by the contractance property shown in 
   Definition~\ref{def:splitting function}, 
   for all $i \in \{1,\ldots,k\}$ $d_{ji} \subset d_j$. 
   Therefore, by 
   Definition~\ref{ordering on stores}, for all $i \in \{1,\ldots,k\}$ 
   $S[d_j/d_{ji}]\prec_s S$. 

\textbf{Case (b).} 
   By Definition~\ref{def:selecting function}, $\#d_j>1$ and, 
   by the completeness 
   property of the splitting functions shown in 
   Definition~\ref{def:splitting function}, 
   \begin{align}
   \label{s in dj}   
       \forall s \in d_j \st \exists i \in \{1,\ldots,k\}: s \in d_{ji}
   \end{align}
   
   Suppose that $S'=(d_1',\ldots,d_n')$. By Definition~\ref{solution}, 
   for all $i \in \{1,\ldots,n\}$ $d_i'=\{s_i'\}$ and also, as $S' \prec_s S$,
   $s_j' \in d_j$. As consequence, by~\eqref{s in dj}, 
   $\exists i \in \{1,\ldots,k\}: s_j' \in d_{ji}$   so that $d_j' \subseteq d_{ji}$. 
   Therefore, by Definition~\ref{ordering on stores},
   $S' \preceq_s S[d_j/d_{ji}]$.    
$\blacksquare$ 
\end{proof}

\noindent {\bf Theorem~\vref{properties of the schema}.} \label{Proof of 
termination} 

\begin{proof} (Property (1). Termination)
In the following, we show that the search tree for $\branch({\cal 
C},S,\p,\varepsilon)$  is finite so that the procedure effectively terminates. 

Let $S_\epsilon = S$ and $\p_\epsilon = \p$. If the search tree for $\branch({\cal 
C},S_\epsilon,\p_\epsilon,\varepsilon)$ has only one node then the procedure 
terminates.  Otherwise, the root node $\epsilon$ has $k$ children with constraint 
stores $S_{i}$ where $i \in \{1,\ldots,k\}$ and  $S_{i}=S_\epsilon^f[d_{j}/d_{ji}]$. 
By Lemma~\ref{lemma:ordering on splitting}(a) and~Definition~\ref{filtering 
function}, for all $i \in \{1,\ldots,k\}$, $S_{i} \prec_s S_\epsilon$ and, by 
Proposition~\ref{precision monotonicity}, $\precision(S_{i}) <_{\RI} 
\precision(S_\epsilon)$. Then, $\precision(S_{i}) <_{\RI} \top_{\RI}$. Suppose now 
that $\precision(S_{i}) = (\top_{\Re},n)$ for some $n \in \Integer$. Then the test 
in Line~\ref{lab:main2} $\p_{i}-\precision(S_{i})\leq (\varepsilon,0)$ holds and the 
node containing $S_{i}$ has no children.  Otherwise, for all $i \in \{1,\ldots,k\}$, 
\begin{align}
\label{eq:properties of the schema:pi-step1} p_i - \precision(S_i) &>_{\RI} 
(\varepsilon, 0)\\ 
\intertext{%
and there exists some constant $\ell \in \Re$ such that } \nonumber \precision(S_i) 
&<_{\RI} (\ell \times \varepsilon,0). 
\end{align}

We show by induction on the length $j\ge 1$ of a path $q$ in the search tree that 
\begin{equation*} 
\precision(S_i) - \precision(S_q^f) \ge_{\RI} 
     \bigl((j-1) \times \varepsilon,0\bigr).
\end{equation*}
It follows that $j \leq \ell$ and that, all paths have length $\leq \ell + 1$ (since 
the second condition in Line~\ref{lab:main3} in Figure~\ref{fig:generic schema} 
holds) and thus there are no infinite branches. 

The base case when $j=1$ follows from~(\ref{eq:properties of the schema:pi-step1}). 
Suppose next that $j>1$ and that the hypothesis holds for a path $q$ of length 
$j-1$. Let $q \concat i_q$ be a child of $q$ of length $j$. Then, by the condition 
in Line~\ref{lab:main3} of the {\it if} sentence, 
\[
\p_{q \concat i_q}-\precision(S_{q \concat i_q}^f) >_{\RI} (\varepsilon,0), 
\]
However, by the inductive hypothesis, 
\begin{align*}
\precision(S_i) - \precision(S_{q}^f) &\geq_{\RI} 
   \bigl((j-2)\times \varepsilon, 0\bigr)
\intertext{%
so that, as $\precision(S_{q}^f	)$ is $\p_{q \concat i_q}$, } \precision(S_i)\! -\! 
\precision(S_{q \concat i_q}) &\geq_{\RI} 
    (\varepsilon,0)+ \bigl((j-2)\times \varepsilon,0\bigr)
    =\bigl((j-1)\times \varepsilon,0\bigr). \blacksquare
\end{align*}
\end{proof}

\begin{proof} (Property (2). Completeness) 
Let $R$ be a solution for $\langle {\cal V}, {\cal D}, {\cal C} \rangle$. Then, $R 
\preceq_s S_\epsilon$ and, by Definition \ref{filtering function}, $R \preceq_s 
S_\epsilon^f$. If  $R = S_\epsilon^f$, then $S_\epsilon^f$ is consistent and not 
divisible by Definition~\ref{solution},  tests in 
Lines~\ref{lab:main2}-\ref{lab:main3} hold and $R$ is pushed in the stack $P$. 
Otherwise, $R\prec_s S_\epsilon^f$ and, by Definition~\ref{solution}, $S_\epsilon^f$ 
is consistent and divisible. As $p_\epsilon=\top_{\RI}$, then condition in 
Line~\ref{lab:main3} does not hold and node $\epsilon$ has $k$ children. By 
Lemma~\ref{lemma:ordering on splitting}(a) and Definition~\ref{filtering function}, 
for any $q$ of length $m\ge1$, $S_{q \concat i_q}^f\prec_s S_{q}^f$ and by  
Proposition~\ref{precision monotonicity},  $\precision(S_{q}^f) - \precision(S_{q 
\concat i_q}^f)>(0.0)$ that means that the condition $\p_{q \concat i_q} - 
\precision(S_{q \concat i_q}^f)\leq(\varepsilon,0)$ in Line~\ref{lab:main3} never 
holds. It follows that all the branches in the tree terminate either with an 
inconsistent store (because test in Line~\ref{lab:main2} does not hold) or with a 
consistent and not divisible store (as result of holding tests in 
Lines~\ref{lab:main2} and ~\ref{lab:main3}). Now, we show by induction on the length 
$j\ge 1$ of a path $q$ in the search tree that 
\begin{align}
\label{completeness equation} R \prec_s S_{q}^f \implies \exists i_q \in 
\{1,\ldots,k\}: R \preceq_s S_{q \concat i_q}^f. 
\end{align}

As, by hypothesis, the procedure terminates then the search tree is finite and it 
follows that there exists some path $p=q \concat q'$ such that $R = S_{p}^f$. Then, 
by Definition~\ref{solution} $S_{p}^f$ is consistent and not divisible and tests in 
Lines~\ref{lab:main2} and~\ref{lab:main3} hold so that $R$ is put in the stack $P$. 
In the base case, when $j=1$,   $S_{i}=S_\epsilon^f[d_j/d_{ji}]$ ($i \in 
\{1,\ldots,k\}$) and by Lemma~\ref{lemma:ordering on splitting}(b) and 
Definition~\ref{filtering function} $\exists i \in \{1,\ldots,k\}: R \preceq_s 
S_{i}^f$. Suppose next that $j>1$ and that the hypothesis holds for a path $q$ of 
length $j-1$ so that $R \preceq_s S_{q}^f$. If $R \prec_s S_{q}^f$ then, by 
Definition~\ref{solution}, $S_{q}^f$ is divisible so that the node $S_{q}^f$ has $k$ 
children by Lemma~\ref{lemma:ordering on splitting}(b)and Definition~\ref{filtering 
function} 
 $\exists i_{q} \in \{1,\ldots,k\}: R 
\preceq_s S_{q \concat i_q}^f$. $\blacksquare$ 
\end{proof}

\begin{proof} (Property (3). Aproximate completeness)
Let $R$ be a solution for $\langle {\cal V}, {\cal D}, {\cal C} \rangle$ so that 
$R\preceq_s S_\epsilon$ and, by Definition~\ref{filtering function},  $R \preceq_s 
S_\epsilon^f$. Since $\varepsilon>0.0$, and as shown in Theorem~\ref{properties of 
the schema}(1), all paths have length $\leq \ell + 1$. Therefore,  as shown in 
completeness proof, by following ~\eqref{completeness equation}, there must exists 
some path $q$ with no children and length $j \ge 1$ such that $R \preceq_s S_{q}^f$. 
If $R = S_{q}^f$ then $R$ is put in the stack since, by Definition~\ref{solution}, 
$R$ is consistent and not divisible and thus tests in Lines \ref{lab:main2} 
and~\ref{lab:main3} hold. Otherwise, as shown in termination proof, the node 
$S_{q}^f$ has no more children since the test $\p_{q}-\precision(S_{q}^f)\leq_{\RI} 
(\varepsilon,0)$ holds and $S_{q}^f$ is put in the stack.  As $R \preceq_s S_{q}^f$, 
by Definition \ref{solution}, $S_{q}^f$ is either a solution for $\langle {\cal V}, 
{\cal D}, {\cal C} \rangle$ or a partial solution that covers $R$. $\blacksquare$ 
\end{proof}

\begin{proof} (Property (4). Correctness)
Let $R$ be an element in the final state of $P$. As shown in completeness proof, if 
$\varepsilon=0.0$ the test $\p_{q}-\precision(S_{q}^f)\leq(\varepsilon,0)$ never 
holds, for all path $q$ of length $m \ge 1$ (also, Line~\ref{lab:main3} is never 
satisfied when $q=\epsilon$). Therefore, $R$ is in $P$ because there exists a path 
$q$ where $S_{q}^f=R$ is consistent and not divisible so that tests in 
Lines~\ref{lab:main2} and~\ref{lab:main3} hold. Moreover, by 
Definition~\ref{filtering function}, $S_{q}^f$ is  a solution for $\langle {\cal V}, 
{\cal D}, {\cal C} \rangle$. $\blacksquare$ 
\end{proof}

\begin{proof} (Property (5). Approximate correctness or control on the
      result precision)
By Theorem~\ref{properties of the schema}(4), if $R \in P_{0.0}$ then $R$ is a 
solution for $\langle {\cal V}, {\cal D}, {\cal  C} \rangle$ and,  by 
Theorem~\ref{properties of the schema}(2), if $R$ is a solution for $\langle {\cal 
V}, {\cal D}, {\cal  C} \rangle$ then $R \in P_{0.0}$. Also, by 
Theorem~\ref{properties of the schema}(3), $\exists S_{\varepsilon_1} \in 
P_{\varepsilon_1}$ and $\exists S_{\varepsilon_2} \in P_{\varepsilon_2}$ such that 
$R \preceq_s S_{\varepsilon_1}$ and $R \preceq_s S_{\varepsilon_2}$. Thus, by 
Definition~\ref{ordering on stacks}, $P_{0.0}\preceq_p P_{\varepsilon_1}$ and 
$P_{0.0}\preceq_p P_{\varepsilon_2}$ (also $P_{\varepsilon_1}$ and 
$P_{\varepsilon_2}$ are not empty). 

Now we prove that $P_{\varepsilon_1} \preceq_p P_{\varepsilon_2}$. Suppose that 
$S_{\varepsilon_2} \in P_{\varepsilon_2}$. Then, exists a path $q$ of length $m \ge 
0$ such that $S_{q}^f=S_{\varepsilon_2}$ and $S_{\varepsilon_2}$ was pushed in the 
stack because tests in Lines~\ref{lab:main2} and~\ref{lab:main3} hold so that 
$S_{\varepsilon_2}$ is consistent and (a) also not divisible (i.e., it is a 
solution), or (b) the test $\p_{q}-\precision(S_{q}^f)\leq(\varepsilon_2,0)$ holds. 

Suppose (a). If $q=\epsilon$ then it is obvious that $S_{\varepsilon_2}$ is also 
pushed in $P_{\varepsilon_1}$. Otherwise, for all path $q_1$ with length $\ge 1$ and 
where $q=q_1 \concat q_2$  the test $\p_{q_1}-\precision(S_{q_1}^f)>(\varepsilon,0)$ 
holds for $\varepsilon=\varepsilon_2$ and thus also holds for 
$\varepsilon=\varepsilon_1$. Therefore, $S_{\varepsilon_2}$ is also pushed in 
$P_{\varepsilon_1}$.  Now suppose (b). Then, if 
$\p_{q}-\precision(S_{q}^f)\leq(\varepsilon_1,0)$ then $S_{\varepsilon_2}$ is also 
pushed in $P_{\varepsilon_1}$, otherwise since  $S_{q}^f$ covers a solution and as 
shown in proof of Theorem~\ref{properties of the schema}(3) there is be some path 
$q'=q \concat p$ such that the test 
$\p_{q'}-\precision(S_{q'}^f)\leq(\varepsilon_1,0)$ holds and thus $S_{q'}^f$ is 
pushed in  $P_{\varepsilon_1}$. By induction on the length of $p$ and 
by~\eqref{completeness equation}, it is straightforward to prove that $S_{q'}^f 
\preceq_s S_{q}^f$. Thus, by Definition~\ref{ordering on stacks}, $P_{\varepsilon_1} 
\preceq_p P_{\varepsilon_2}$. $\blacksquare$ 
\end{proof}

\noindent {\bf Theorem~\vref{properties of the extended schema}.} 

\begin{proof} (Property (1). Termination)
This proof is as that of Theorem~\ref{properties of the schema}(1). $\blacksquare$ 
\end{proof}

\begin{proof} (Property (2))
Observe that if $\fcost(S)=\delta$ for all $S \in {\cal D}$, then test in 
Line~\ref{lab:main4*} of the extended schema always holds. It is straightforward to 
prove, in this case, that the schemas $\branch/4$ and $\branch_+/4$ are equivalent 
so that all properties for the schema $\branch/4$ are also held in the schema 
$\branch_+/4$. $\blacksquare$ 
\end{proof}

\begin{proof} (Property (3). Soundness on optimisation)
We prove the case when $\diamond$ and $\delta$ are, respectively, $>$ and 
$\bot_{D_<}$. The respective case is proved analogously. As shown in proof of 
Theorem~\ref{properties of the schema}(2), for $\varepsilon=0.0$, if $R$ is a 
solution for $\langle {\cal V}, {\cal D}, {\cal C} \rangle$ then there exists some 
path $q$ of length $j\ge 0$, such that \(R=S_{q}^f \) and the tests in 
Lines~\ref{lab:main2}-~\ref{lab:main3} hold by Definition~\ref{solution}. Thus,  
Line~\ref{lab:main4*} is reached for each solution  $R \in \Sol(\langle {\cal V}, 
{\cal D}, {\cal C} \rangle)$, and as consequence, the top of $P$ will contain the 
first solution found that maximises $\fcost/1$. $\blacksquare$ 
\end{proof}

\noindent {\bf Theorem~\vref{more properties on optimisation}.} 

\begin{proof} (Property: Approximate soundness)
We show that during the execution of $\branch_+({\cal C},S,\p,\varepsilon_1)$, 
Line~\ref{lab:main4*} is reached for some $S_{q}^f \preceq_s top(P_{\varepsilon_2})$ 
(where $q$ is a path of length $m_1 \ge 0$). As consequence, $\fcost(S_{q}^f) 
\preceq_{D_<} \fcost(\topp(P_{\varepsilon_2}))$. It follows that either $S_{q}^f$ is 
in the top of $P_{\varepsilon_1}$ or $S_{q}^f$ is not in the top of 
$P_{\varepsilon_1}$ since $\fcost(\topp(P_{\varepsilon_1})) \preceq_{D_<} 
\fcost(S_{q}^f)$ so that effectively $\fcost(\topp(P_{\varepsilon_1})) \preceq_{D_<} 
\fcost(\topp(P_{\varepsilon_2}))$. 

Observe that $\topp(P_{\varepsilon_2})$ is in $P_{\varepsilon_2}$ because there 
exist some path $q'$ with length $m_2 \ge 0$ such that 
$S_{q'}^f=\topp(P_{\varepsilon_2})$ and tests in 
Lines~\ref{lab:main2}-\ref{lab:main4*} hold. Then, as shown in proof of 
Theorem~\ref{properties of the schema}(1), for all path $q_1$ such that $q'=q_1 
\concat q_2$ 
\[
   \p_{q_1}-\precision(S_{q_1}^f)>(\varepsilon_2,0).
\]

As $\varepsilon_1 < \varepsilon_2$, this also holds in the execution of 
$\branch_+({\cal C},S,\p,\varepsilon_1)$. Therefore, if $S_{q'}^f$ is in 
$P_{\varepsilon_2}$ because it is consistent and not divisible, then in the 
execution of $\branch_+({\cal C},S,S',\varepsilon_1)$ Line~\ref{lab:main4*} is also 
reached with $S_{q'}^f$. Otherwise, $S_{q'}^f$ is in $P_{\varepsilon_2}$ because \( 
   \p_{q'}-\precision(S_{q'}^f)\leq(\varepsilon_2,0).
\). Then, as  $\varepsilon_1 < \varepsilon_2$, and as shown in proof of 
Theorem~\ref{properties of the schema}(5), there exists some path $q''$ with length 
$r \ge m_2$ such that 
\[
   \p_{q''}-\precision(S_{q''}^f)\leq(\varepsilon_1,0)\text{ and }S_{q''}^f \preceq_s S_{q'}^f.
\]
so that again Line~\ref{lab:main4*} is reached for $S_{q''}^f$. 
\end{proof}

\newpage

\section*{Appendix: A simple example}

Here we show a simple example in the domain $\Interv(Integer)$, illustrating the 
flexibility of the schema to solve a CSP in different ways. Let $\langle {\cal V}, 
{\cal D}, {\cal C} \rangle$ be a CSP where ${\cal V}=\{x_1,x_2,x_3\}$, 
$\Doms=\{\Interv(\Integer)\}$ and ${\cal C}$ is the constraint set 
\[
\{ x_1 + x_2 + x_3 \leq 1,\ x_1 \leq 1, \ x_2 \leq 1,\ x_3 \leq 1, 
  \ x_1 \ge 0, \ x_2 \ge 0,\ x_2 \ge 0\}. 
\]
Consider also the following cost functions\footnote{The sum is defined to return the 
mid point of the sum of operand intervals, e.g., the sum of two intervals a..b and  
c..d is (a+c+b+d)/2 that is exactly the mid point of the interval a+c..b+d.} defined 
on different ranges: 
\begin{alignat*}{2}
&\fcost_1(x_1,x_2,x_3)=1.0.                              & \qquad\text{Range: }\Re\  
\\ &\fcost_2(x_1,x_2,x_3)=x_1+x_2+x_3.                      & \qquad\text{Range: 
}\Re\  \\ &\fcost_3(x_1,x_2,x_3)=(\fcost_2(x_1,x_2,x_3),x_1+x_3).  & 
\qquad\text{Range: }\Re^2 \\ &\fcost_4(x_1,x_2,x_3)=(\fcost_2(x_1,x_2,x_3),x_2+x_3).  
& \qquad\text{Range: }\Re^2 
\end{alignat*}

Consider now the instance $\branch_{\Interv(FD)}$, as defined in Section~\ref{close 
FD interval instances}, where $FD=\Integer$ and assume that $\ch_{\naive}$ is as 
defined  in Example~\ref{ex:naive labeling}, $\p=\top_{\RI}$, $\varepsilon=0.0$ and 
initially the global stack $P$ is empty. Suppose that as filtering algorithm we 
define a simple consistency check on the consistency of constraint stores in such a 
way that $\filter({\cal C},S)$ returns $S$ if $S$ is consistent and the inconsistent  
store $(\emptyset,\emptyset,\emptyset)$ otherwise. Now, assume that $\branch_+({\cal 
C},S,p,\varepsilon)$ is executed with different values for $\delta$, $\diamond$ and 
$\fcost/1$. Since the domain is finite, termination is guaranteed even if 
$\varepsilon=0.0$. Each row in Table~\ref{example instances} corresponds with a 
different execution of the extended schema where 
\begin{itemize}
\item Column 1 indicates the way in which the CSP is solved,
\item Column 2 shows the value to which $\delta$ is initialised, 
\item Column 3 the cost function used in the current instance, 
\item Column 4 the initialisation of $\diamond$, 
\item Column 5 indicates where is, in the global stack, the solution(s) and 
\item Column 6 references the figure that shows the final state of the stack 
$P$\footnote{To the right of each element in $P$ we write its cost.}. 
\end{itemize}
\begin{table}[thb]
\footnotesize \centerline{ 
\begin{tabular}{|l|c|c|c|c|c|}    \hline
CSP Type         &  $\delta$                        & cost function & $\ \diamond\ $ 
& Solution      & Figure \\ \hline Classical CSP    &    1.0                           
&      $\fcost_1$     &       $=$      & Each element in the stack & 
\ref{optimisation stacks}(a) \\ Maximisation COP &  $\bot_{\Re}$                 &      
$\fcost_2$     &       $>$      & Stack Top    & \ref{optimisation stacks}(b) 
\\ Minimisation COP &  $\top_{\Re}$                  &      $\fcost_2$     &       
$<$      & Stack Top    & \ref{optimisation stacks}(c) \\ Max-Min COP (i)  &  
$(\bot_{\Re},\top_{\Re})$ &      $\fcost_3$     &       $<_1$    & Stack Top    & 
\ref{optimisation stacks}(d) \\ Max-Min COP (ii)  &  $(\bot_{\Re},\top_{\Re})$ &      
$\fcost_4$     &       $<_1$    & Stack Top    & \ref{optimisation stacks}(e) 
\\ Max-Min COP (iii)&  $(\bot_{\Re},\top_{\Re})$ &      $\fcost_3$     &       
$<_2$    & Stack Top    & \ref{optimisation stacks}(f) \\ Max-Min COP (iv) &  
$(\bot_{\Re},\top_{\Re})$ &      $\fcost_4$     &       $<_2$    & Stack Top    & 
\ref{optimisation stacks}(g) \\ 
 \hline 
\end{tabular}
} \caption{Different solvings of the CSP} \label{example instances} 
\end{table}                                                                         
\normalsize 
\begin{table}[htb]
\footnotesize \centerline{ 
\begin{tabular}{|c|c|c|c|c|}    \hline
Solution (S) & $\fcost_1(S)$ & $\fcost_2(S)$ & $\fcost_3(S)$ & $\fcost_4(S)$ 
\\ \hline 
  (1,0,0)    &     1.0       &     1.0       &   (1.0,1.0)   &   (1.0,0.0)  \\
  (0,1,0)    &     1.0       &     1.0       &   (1.0,0.0)   &   (1.0,1.0)  \\
  (0,0,1)    &     1.0       &     1.0       &   (1.0,1.0)   &   (1.0,1.0)  \\
  (0,0,0)    &     1.0       &     0.0       &   (0.0,0.0)   &   (0.0,0.0)  \\ \hline 
\end{tabular}
} \caption{Evaluation of the solutions to the problems} \label{evaluation of 
solutions} 
\end{table}                                                                         
%
\begin{figure}[htb]
\centerline{ \epsfysize 4.0cm \epsfbox{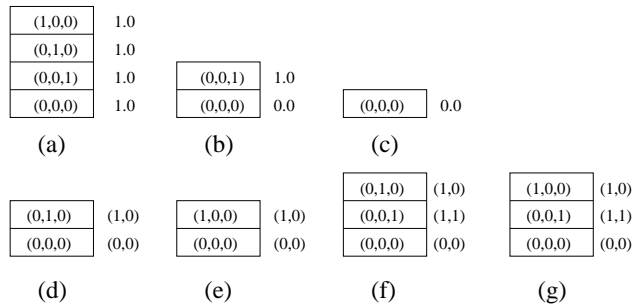}} \caption{\footnotesize The final 
state of the global stack P in the different solvings of the CSP} 
\label{optimisation stacks} 
\end{figure}
By simplicity, suppose that during each execution of the extended schema,  branches 
are solved by classical backtracking following a classical depth first strategy. 
Then, the  CSP is solved in different ways. For instance, to solve the problem as a 
classical CSP (see Row 1 in Table~\ref{example instances}), $\fcost$ is a constant 
function with value $\delta$ (where $\delta$ is 1.0) and $\diamond$ is $=$. Then, 
all possible solutions for the problem are pushed in the stack (see 
Figure~\ref{optimisation stacks}(a)).  

Also, Rows 2-3 in Table~\ref{example instances} show how to solve this CSP by 
maximising and minimising the function $\fcost_2$ respectively. The optimal solution 
is that on the top of the stack (see Figures~\ref{optimisation stacks}(b) 
and~\ref{optimisation stacks}(c)). On their turn, Rows 4-7 indicate how to mix 
optimisation criteria to solve the CSP. For instance, assume the following two 
orderings on $\Re^2$: 
\begin{align*}
(a,b) \leq_1 (c,d) \iff a \ge c \text{ and } b \leq d; \\ (a,b) \leq_2 (c,d) \iff a 
> c \text{ or } a = c \text{ and }b \leq d. 
\end{align*}           
Then, row 4 corresponds to the problem of maximising $x_1+x_2+x_3$ and minimising 
$x_1+x_3$ whereas row 5 corresponds to the problem of maximising $x_1+x_2+x_3$ and 
minimising $x_2+x_3$. Also, row 6 corresponds to the problem of firstly maximising 
$x_1+x_2+x_3$, and if this cannot be more optimised then minimise $x_1+x_3$ (this is 
consequence of the ordering $<_2$) whereas row 7 does the same but minimising 
$x_2+x_3$. Figure~\ref{optimisation stacks} shows the final state of the global 
stack for each of these cases (also Table~\ref{evaluation of solutions} shows the 
evaluation of each solution to the CSP by the different cost functions
\end{document}